\documentstyle[amsmath,amssymb,graphicx]{article}

\def\be{\begin{eqnarray}}
\def\ee{\end{eqnarray}}
\def\nn{\nonumber}

\def\tr{{\rm tr}\,}

\hoffset= - 1.0in         

\begin{document}

\hfill ITEP/TH-16/12

\bigskip

\centerline{\Large{Faces of matrix models
}}

\bigskip

\centerline{A.Morozov}

\bigskip

\centerline{\it ITEP, Moscow, Russia}

\bigskip

\centerline{ABSTRACT}

\bigskip

{\footnotesize
Partition functions of eigenvalue matrix models
possess a number of very different descriptions:
as matrix integrals, as solutions to linear
and non-linear equations,
as $\tau$-functions of integrable hierarchies
and as special-geometry prepotentials,
as result of the action of $W$-operators and
of various recursions on elementary input data,
as gluing of certain elementary building blocks.
All this explains the central role of such matrix models in
modern mathematical physics:
they provide the basic "special functions"
to express the answers and relations between them,
and they serve as a dream model of what one should try
to achieve in any other field.

}

\bigskip

\bigskip


Matrix model theory \cite{MaMo} studies the integral
\be
Z_N(g) \sim \int_{N\times N} dM\, e^{-\frac{1}{2g}\tr M^2}
= \int \prod_{i=1}^N dM_{ii} \prod_{i<j}^N d^2M_{ij} \exp \left\{-\frac{1}{2g}
\left(\sum_{i=1}^N M_{ii}^2 + 2\sum_{i<j}^N|M_{ij}|^2\right)\right\}
\label{gaidef}
\ee
over $N\times N$ Hermitian matrices $M$
as a toy-example of quantum field and even string theory.
It is spectacular, how much one can learn from this
seemingly obvious problem.

\bigskip

What does it mean to study an integral?

\bigskip

{\bf First,} we can simply take it.
In this particular case the answer is simple:
\be
Z_N(g) \sim (2\pi g)^{N^2/2}
\label{gaians}
\ee
and does not look very interesting.
However it only seems so.
As usual, of interest is not the answer itself, but its
{\it decomposition}, implied, by internal structure
of our "theory".
And the more we know about these structures.
the more interesting decompositions we can obtain.
In this particular case we could notice that
$M = UDU^\dagger$, where $D = {\rm diag}\{x_i\}$
matrix, made from eigenvalues of $M$, and $U$
is a unitary matrix.
Then the same integral is decomposed into two --
over unitary matrix $U$ and over $N$ eigenvalues $\{x_i\}$.
Factoring away the volume $V_N$ of the unitary group, we obtain:
\be
Z_N =
\frac{1}{N!} \int  \prod_{i<j}^N (x_i-x_j)^2 \prod_{i=1}^M e^{-x_i^2/2g} dx_i
= \frac{V_1^N}{N!V_N}\int_{N\times N} dM e^{-\frac{1}{2g}\tr M^2}
\label{ZN}
\ee
This is already a somewhat non-trivial decomposition, because
\be
V_N = \frac{(2\pi)^{N(N+1)/2}}{\prod_{k=1}^N k!}
\ee
what is a considerably more complicated expression than the original (\ref{gaians}).

\bigskip

{\bf Second,} to study an integral in QFT sense means to treat it as measure,
and consider all possible {\it correlators}.
This means that  of interest is not the (\ref{gaidef}) itself, but the averages
\be
C_{i_1,\ldots,i_k} = \Big< \tr M^{i_1}\ldots \tr M^{i_k} \Big> =
\frac{\int \tr M^{i_1}\ldots \tr M^{i_k} e^{-\frac{1}{2g}\tr M^2} dM}
{\int e^{-\frac{1}{2g}\tr M^2} dM}
\ee
or even their connected counterparts, like
\be
C^{conn}_{ij} = C_{ij} - C_iC_j
\ee
This is already a far-less-trivial problem, and looking at the very first
examples one immediately observes an emergency of new structure:
\be
C_0(N) = N \nn\\
C_2(N) = gN^2 \nn\\
C_4(N) = g^2(2N^3+N) \sim 2(gN)^3 + g^2(gN)\nn\\
C_6(N) = g^3(5N^4+2N^2) \sim 5(gN)^4 + 2g^2(gN)^2 \nn\\
\ldots
\ee
The fact that each correlator is a polynomial (not a monomial) in $N$
is encoded in the idea of {\it loop expansion}.
The fact that all coefficients are integers signals about connection
to combinatorics and is encoded in the idea of {\it topological theories}.

\bigskip

{\bf Third,} if we move in the direction of string theory,
we need not just correlators: we need {\it generating functions}.
For the set of $C_{i_1,\ldots, i_k}$ there are two obvious options:
\be
Z\{t\} = \frac{V_1^N}{N!V_N}
\int dM e^{-\frac{1}{2g}\tr M^2 + \sum_{k=0}^\infty t_k\tr M^k}
= e^{F\{t\}}
\label{pft}
\ee
and
\be
\rho^{(m)}\{z\} = \left< \prod_{i=1}^m \tr\frac{dz_i}{z_i-M} \right>
\label{resodef}
\ee
Then we have
\be
C_{i_1\ldots i_k} = \frac{1}{Z_N}\frac{\partial^k Z_N}{\partial t_{i_1}\ldots\partial t_{i_k}}, \nn\\
C_{i_1\ldots i_k}^{conn} = \frac{\partial^k \log Z_N}{\partial t_{i_1}\ldots\partial t_{i_k}}
\ee
and
\be
\rho^{(m)}\{z\} = \frac{1}{Z_N}\prod_i^m \hat\nabla(z_i) Z_N
\ee
where $\hat\nabla(z) = \sum_{k=0}^\infty \frac{dz}{z^{k+1}}\frac{\partial}{\partial t_k}$.
One can also introduce the connected resolvent
\be
\rho_{conn}^{(m)}\{z\} = \prod_i^m \hat\nabla(z_i) \log Z_N
\ee
The fact that correlators $C_I$ where polynomials in $N$ is now expressed in the
{\it genus expansion} of the free energy and connected resolvents:
\be
F\{t|g,N\} = \sum_{p=0}^\infty g^{2p-2} F_p\{t|gN\}
\ee
and similarly
\be
\rho^{(m)}_{conn}\{z\} = \sum_{p=0}^\infty g^{2p} \rho^{(\!p\,|m)}\{z|gN\}
\ee
Already at this stage something highly non-trivial shows up.
This becomes clear from a look on the first few resolvents:
\be
\rho^{(0|1)}(z) = \frac{y(z)dz}{2} \nn \\ \nn\\
\rho^{(1|1)}(z) = \frac{dz}{y^5(z)} \nn \\ \nn \\
\rho^{(0|2)}(z_1,z_2) = \frac{1}{(z_1-z_2)^2}\frac{dz_1dz_2}{y(z_1)y(z_2)}\nn \\
\ldots
\ee
They all are meromorphic (poly)differentials on
a Riemann surface
\be
\Sigma: \ \ \ y^2 = z^2 - 4(gN)
\ee
which is called the {\it spectral curve}.

\bigskip


According to the string-theory approach, from this point
we should move far enough in a number of different directions.

\bigskip

{\bf  Other phases.}
As soon as we introduced the generating function $Z\{t\}$,
we can start treating it non-perturbatively.
This means that $t_k$ are considered not just as infinitesimal expansion
parameters, defining a {\it germe}, but as the coupling constants,
and study what happens when they take finite (or even infinite) values.
Then $Z\{t\}$ defines a partition function of a {\it family}
of theories, called {\it non-perturbative} partition function.
This partition function can be re-expanded not only around the
Gaussian point, but around any {\it background potential}
$V(M) = \sum_k T_k M^k$. Partition function (particular branch of it)
then becomes also a function of parameters $T_k$, which parameterize the
{\it moduli} of the spectral curve. Phase transitions take place when the genus
of the curve changes -- it is controlled by the number of extrema
of the background potential. The study of these dependencies is
the subject of {\it Seiberg-Witten theory} \cite{SW}, in matrix-model context
the corresponding field is sometime called the theory
multi-cut solutions  or of the {\it Dijkgraaf-Vafa phases} \cite{DVph}.
The particular {\it branch} of partition function is also known as
{\it CIV prepotential} \cite{CIV}.
The most interesting feature of this prepotential are Seiberg-Witten
special-geometry
equations, describing dependence on the moduli by introducing very special
"{\it flat}" coordinates $a_k$ instead of $T_k$:
\be
\left\{\begin{array}{c}
a_k = \oint_{A_k} \Omega  \\ \\
\frac{\partial F}{\partial a_k} = \oint_{B_k}\Omega
\end{array}\right.
\ee
and the role of the Seiberg-Witten differential on the spectral curve
is presumably played by the 1-point resolvent $\Omega(z) = \rho^{(1)}(z)$
\cite{towaproof}. The system of interrelated multidensities $\rho^{(p|m)}_\Sigma$
can in fact be built in a universal way for arbitrary Seiberg-Witten
family of spectral curves $\Sigma$ -- this procedure is now known as
AMM/EO topological recursion \cite{AMM/EO} and has surprisingly many applications.
Whenever partition function can be reconstructed in this way, this signals
about the matrix-model hidden behind the scene -- and there are already
numerous examples, when recursion works, but the matrix model is not yet found.

\bigskip

{\bf Various limits.}
Non-perturbative partition function has a huge variety of different limits
and critical behaviors in the vicinities of all its numerous singularities.
The standard large-$N$, genus-zero and multiscaling limits are just the examples.
Related problem is the study of convergency properties of various
perturbative series. All this is very important in applications
and constitutes, perhaps, the biggest parts of traditional matrix-model theory.

\bigskip

{\bf Other observables.}
In string-theory paradigm there is no special preference for any {\it obvious}
choice of observables. Instead of the correlators $C_I$
of the monomials $\tr M_i$ one could study those, say, of the "Wilson loops"
$\tr\! \left(e^{sM}\right)$,
and form many other generating functions, different from (\ref{pft}) and (\ref{resodef}),
like the celebrated Harer-Zagier exact 1-point function \cite{HZ}
\be
\phi(z|\lambda) = \sum_{N=0}^\infty \lambda^N
\sum_{k=0}^\infty \frac{z^{2k}<\tr M^{2k}>}{(2k-1)!!}
= \frac{\lambda}{(1-\lambda)\Big((1-\lambda)-(1+\lambda)z^2\Big)}
\ee
and Brezin-Hikami integrals \cite{BH}
\be
\left< \prod_{i=1}^k \tr\!\left( e^{s_iM}\right)\right>\ = \
\prod_{i=1}^k \frac{e^{s_i^2/2}}{s_i}\oint e^{u_is_i}du_i \left(1+\frac{s_i}{u_i}\right)^N
\prod_{i<j}^k\frac{(u_i-u_j)(u_i-u_j+s_i-s_j)}{(u_i-u_j+s_i)(u_i+u_j-s_j)}
\label{BH}
\ee
The number of integrals here is $k$, not $N$, as in (\ref{ZN}).
In fact, these two subjects are unexpectedly closely related \cite{HZBH}.
Harer-Zagier functions capture contributions from all genera --
they differ from (\ref{pft}) by a kind of Pade transform and allow to
put under the control the divergence of perturbative genus expansion.
Instead they hide all the information related to spectral curves
and Seiberg-Witten equations -- but are capable to provide a closed expression for the
Seiberg-Witten differential $\Omega(z) = \rho^{(1)}(z)$.
Unfortunately, they are much more difficult to study than the resolvents.

\bigskip

{\bf Alternative formulations.}
For non-perturbative partition functions integrals (be they matrix
or functional) provide only a description of particular phases:
or, in worst case just the perturbative {\it germes} at particular points.
More adequate are formulations in terms of {\it $D$-modules}
or $\tau$-functions,
characterizing partition functions as solutions to linear
or quadratic equations respectively.
It is still unclear, how general is the existence of quadratic
(integrability-theory) structures and
if higher non-linearities can also be relevant.
At this moment, the "{\it matrix-model $\tau$-functions}" --
usually, KP/Toda-functions, satisfying also a linear {\it string equation},
and, as a corollary, a whole infinite set of linear "Virasoro constraints"
\cite{virco}
are the most profound special functions, encountered in modern
mathematical physics. They are natural for presentation of quantitative
results in various fields of string theory, and their investigation is one
of the primary purposes of modern science.

\bigskip

{\bf Integrability and $W$-representation.}
Emergency of non-linear (integrable) relations, like \cite{GMMMO}
\be
\frac{\partial^2 \log Z_N}{\partial t_1^2} =
\frac{Z_{N+1}Z_{N-1}}{Z_N^2}
\ee
for (\ref{pft}), is so non-trivial and so universal
in string theory, that it can be considered as one of the main features
of non-perturbative physics -- still very mysterious.
One should look for adequate ways to characterize these structures.
Non-trivial $\tau$-functions can be made from the "trivial" ones by
integrability-preserving transforms, described in terms of the
$W$-operators, which move the points in the Universal Grassmannian,
parameterizing the space of the KP/Toda (free-fermion) $\tau$-functions.
In other words, a matrix-model $\tau$-function can be considered
as a result of the "evolution", driven by cut-and-join ($W$) operators from
some simple "initial conditions" \cite{Wreps}:
\be
Z\{t\} = e^{\hat W} \tau_0\{t\}
\ee
For (\ref{pft}) this $W$-representation looks as follows:
\be
Z_N\{t\} = \exp\left\{\sum_{a,b} \left(at_abt_b\frac{\partial}{\partial t_{a+b-2}}
+ (a+b+2)t_{a+b+2}\frac{\partial^2}{\partial t_a\partial t_b}\right)\right\} e^{Nt_0}
\ee

\bigskip

{\bf Generalizations.}
According to string-theory paradigm, one should not just embed original model in
a set the similar ones by exponentiating all naive observables, one should also
deform everything else, including the discrete parameters.
In application to matrix models this means that starting from (\ref{gaidef})
one should not just switch from quadratic to arbitrary potential, not just treat $N$
as one of parameters, but also substitute Hermitian matrices by others:
unitary, orthogonal, symplectic, belonging to exceptional and other Lie
algebras, to generic tensorial categories etc etc.
Of all this the most far-going so far are extensions to
unitary matrix models \cite{unimo} and to $\beta$-ensembles \cite{betadefo}.
In all cases one expects to find all the relevant representations:
not only through traditional integral formulas, but also
as $D$-modules, as $\tau$-functions, through $W$-operators, through topological
recursion which start from peculiar spectral curves, through Harer-Zagier-type
recursions. Some results in these directions exist, but they are far from being
exhaustive.

\bigskip

{\bf External fields and dualities.}
Another generalization is inclusion of external fields.
The simplest possibility is to switch from (\ref{gaidef}) to
\be
{\cal Z}(\nu|A) = e^{-\frac{g}{2}\tr A^2} \int_{n\times n} e^{-\frac{1}{2g}\tr M^2
+ \tr MA} (\det M)^\nu dM
\label{komdef}
\ee
Determinant is introduced here to make the dependence on $A$ non-trivial,
and we also changed the notation for the size of the integration matrix.
This is done on purpose, because if this function is considered as
a function of the variable $p_k = \tr A^{-k}$, it is actually independent of $n$.
In this way one defines {\it Kontsevich matrix models}, eq.(\ref{komdef})
is the Gaussian one, for properties of generic Kontsevich models see \cite{GKM}.
Really remarkable is the duality between  (\ref{komdef}) and (\ref{pft}):
\be
{\cal Z}(N|A) \sim  Z_N\{t_k\}
\label{dua}
\ee
provided $t_k = \frac{1}{k}p_k = \frac{1}{k}\tr A^{-k}$.
In fact, this duality  \cite{dua} can be used in the derivation of Brezin-Hikami
formulas (\ref{BH}), which, in turn have non-trivial generalization \cite{Okou,HZBH} to
at least the cubic Kontsevich model.

\bigskip

{\bf Unification.}
Duality between Gaussian Hermitian and Kontsevich models is just an example of
interrelation between two {\it a priori} different matrix models.
The goal of string theory is to unify in a similar way all quantum field
theories, and in particular, this applies to unification of all matrix models.
Unification does not mean {\it solving} -- that problem belongs to the field
of non-linear algebra \cite{nal}, which studies formulas like
\be
\int\!\!\int dx dy\ e^{ax^2+bxy +dy^2} \sim \frac{1}{\sqrt{4ad-b^2}} = D_{2|2}^{-1/2}\nn\\
\int\!\!\int  dx dy\ e^{ax^3+bx^2y+cxy^2 +dy^3} \sim {D_{2|3}^{-1/6}}\nn \\
D_{2|3} = 27a^2d^2-b^2c^2-18abcd+4ac^3+4b^3d
\ee
(in general ordinary discriminants $D_{N|r}$ control singularities of integral discriminants).
Unification means that all seemingly different non-perturbative
partition functions either are interrelated (by dualities), or are all reductions
of some larger partition function (arise at particular loci in the extended
space of time-variables), or are all composed from some elementary building blocks.
It turns out that the last, most promising, possibility can be true,
at least in the world of the eigenvalue matrix models.
Namely, at least all the Dijkgraaf-Vafa partition functions can be obtained
by a universal gluing procedure from  a few basic elements \cite{Mth}:
\be
Z\{t\} = e^{\hat U} \prod_{i=1}^k Z^{(i)}\{t^{(i)}\}
\ee
where $\hat U$ is bilinear in derivatives over $t^{(i)}$-variables.
This gluing procedure is closely related to AMM/EO recursion \cite{AMM/EO}
and can be considered as one of its most profound implications.
The role of the elementary building blocks $Z^{(i)}$ play several important
matrix models which posses a sphere with punctures as their spectral curves:
the Gaussian Hermitian model, the cubic Kontsevich model and
the Brezin-Gross-Witten model \cite{BGW}.

\bigskip

\bigskip

{\bf Applications.}
Matrix model theory has infinitely many applications in all branches of science,
far beyond pure mathematics, string theory and even physics.
Still, it deserves mentioning a few relatively new examples, concerning
the abstract fields of research, in order to illustrate once again  the influence
of matrix model intuition on our understanding of basic problems.
These recent applications also emphasize the role of the {\it character calculus}
-- one of the most important matrix-model-theory technical methods.
Moreover, matrix models themselves are not present very explicitly,
what are discovered are the typical structures and relations,
pertinent for matrix-model partition functions.

\bigskip

The first subject is {\bf Hurwitz theory} \cite{Hut}. Today it is clear
that this is basically the story about the algebra of cut-and-join
operators, which are well known in matrix model theory
\be
\hat W_R = \ :\prod_i \tr\left(M\frac{\partial}{\partial M^{tr}}\right)^{r_i}:
\ee
They are labeled by Young diagrams $R = \{r_1\geq r_2\geq\}$
and have Schur functions (the $GL(\infty)$ characters) $\chi_Q[M]$
as common eigenfunctions:
\be
\hat W_R \chi_Q[M] = \varphi_Q(R) \chi_Q[M]
\ee
while eigenvalues $\varphi_Q(R)$ depend on a pair of Young diagrams and
are essentially the characters of symmetric group $S(\infty)$.
The Hurwitz partition functions describe the sums like
\be
\sum_Q d_Q^{2-2p}\varphi_Q(R_1)\ldots\varphi_Q(R_k)
\ \longrightarrow \
\sum_Q d_Q^2 \exp\left(\sum_R t_R\varphi_Q(R)\right)
\ee
and possess many properties, typical for matrix-model $\tau$-functions,
including the deeply hidden Virasoro-constraints,
as well as numerous non-trivial generalizations, involving
non-commutative "open-string" algebra, extending the
commutative "closed-string" one formed by the $\hat W_R$.
See \cite{MMN} for details and references.

\bigskip

The second subject is the {\bf AGT conjecture} \cite{AGT}:
the celebrated identity between $2d$ conformal blocks and
Nekrasov expansions \cite{Nekf} of the  LMNS functions \cite{LMNS},
describing instanton expansions of $4/5/6d$ SYM theories.
This subject brings together conformal field theory,
Seiberg-Witten theory, classical and quantum integrable systems
\cite{intSW}.
At the core of the story is the special "conformal"
matrix model \cite{confmamo}, which realizes Dotsenko-Fateev
representation of conformal blocks, and LMNS functions
appear from Selberg integrals, arising in the character expansion
of the model.
In this language the AGT relation reduces to the Hubbard-Stratanovich
duality \cite{HS}, this works perfectly at $\beta=1$, but
generalization to $\beta$-ensembles remains subtle \cite{towaproof}.
What is extremely important in this story is that the
averages of characters are again characters -- and matrix models
with this special property seem to become more and more
distinguished in modern applications.

\bigskip

The third example is the modern
{\bf theory of knots} \cite{knth,CS}, which studies extended  \cite{extHOMFLY}
HOMFLY $H_R^{\cal B}\{ p_k|\,q\}$ \cite{HOMFLY}
and superpolynomials  $P_R^{\cal B}\{p_k|\,q,t\}$ \cite{superpols}.
A very interesting matrix model realization here is long known for
the underlying Chern-Simons theory, but its generalization
in the presence of non-trivial knots is so far available only
for torus knots \cite{BEM} and for $t=q$.
In general one expects that the model exists, the measure
depends on the braid realization ${\cal B}$, and the HOMFLY
polynomial in representation $R$ is an average of the $SL(N)$ character:
\be
H_R^{\cal B} = \Big< \chi_R[U]\Big>_{\cal B}
\ee
Like in the case of AGT relation, one expects that with this
measure the averages of characters will be again simply re-expanded
in characters, and such model will be a useful tool to study
the character expansions of HOMFLY and superpolynomials,
which are responsible for the fast progress in the field in recent
months.
This is indeed the case for the torus knot $[m,n]$: the measure is given by \cite{BEM}
\be
\Big<\ldots\Big>_{]m,n]} = \ \prod_{i=1}^N \int  e^{-\frac{u_i^2}{mng}}\ du_i\
\prod_{i<j}^N\ \sinh\frac{u_i-u_j}{m}\ \sinh\frac{u_i-u_j}{n}\ \Big(\ \ldots\
 \Big)
\ee
and $\ <\chi_R[U]>_{[m.n]}\  \sim \ \chi_R\left\{kt_k=\frac{[kN]_q}{[k]_q}\right\}$,
moreover, like with all Selberg-type integrals,
this property persists for bilinear combinations of characters.

\bigskip

The forth example, which deserves mentioning is a very similar Chern-Simons type
matrix-model representation
in the very important {\bf ABJM theory} \cite{ABJM}, describing
$N$ copies of $M2$ branes. The only additional
complication is that $\cosh$ factors are also present in denominators
of the Vandermonde determinants. Despite this complication the  model
was completely solved in \cite{ABJMsol} at vanishing times, and the required
non-trivial behavior $\sim N^{3/2}$ (instead of the usual $\sim N^2$)
of the free energy was reproduced in the large-$N$ limit.

\bigskip

Note that adequate introduction of time variables, suitable for
revealing the linear and non-linear relations -- in the form of
Virasoro constraints and KP/Toda integrability respectively --
remains a largely unsolved problem in all these examples,
despite there are already many signals, that these or very similar
structures should exist.
It is one of the primary tasks of matrix-model theory to study and
resolve these mysteries.

\section*{Acknowledgements}

I appreciate the hospitality of OIST, Japan during a workshop,
for which this brief report was originally prepared.
My work is partly supported by Ministry of Education and Science of
the Russian Federation under contract 14.740.11.0081 , by NSh-3349.2012.2,
by RFBR grant  10-01-00536 and
by the joint grants 11-02-90453-Ukr, 12-02-91000-ANF,
11-01-92612-Royal Society and 12-02-92108-Yaf-a.


\begin{thebibliography}{12}

\bibitem{MaMo}
We present here only the list of papers, where additional details can be found
on particular non-traditional subjects, mentioned in the text.
No special references are given to the main papers, which shaped the field
of matrix models.
For basic information and references of this kind see the textbooks:\\ \\
Ph.Di Francesco, P. Ginsparg and J.Zinn-Justin,
Phys.Repts, 254 (1995) 1; \\
M. L. Mehta, Random matrices, Pure and Applied Mathematics Series 142 (2004); \\
{\it The Oxford Handbook of Random Matrix Theory},
Eds. G.Akemann, J.Baik, and Ph.Di Francesco, 2011\\ \\
and, for more specific references, the reviews \\ \\
A.Morozov, Phys.Usp.35 (1992) 671-714; 37 (1994) 1, hep-th/9303139; hep-th/9502091; hepth/
0502010; \\
A.Mironov, Int.J.Mod.Phys. A9 (1994) 4355, hep-th/9312212; Phys.Part.Nucl. 33 (2002) 537; hepth/
9409190; Theor.Math.Phys. 114 (1998) 127, q-alg/9711006



\bibitem{SW}
N.Seiberg and E.Witten, N.Seiberg and E.Witten,
Nucl.Phys. B431 (1994) 484-550, hep-th/9407087;
Nucl.Phys. B426 (1994) 19-52, hep-th/9408099; \\
A.Hanany and Y.Oz, Nucl.Phys., B452 (1995) 283-312, hep-th/9505075;\\
P.Argyres and A.Shapere, Nucl.Phys., B461 (1996) 437-459, hep-th/9509175;\\
J.Sonnenschein, S.Theisen and S.Yankielowicz, Phys.Lett., B367 (1996) 145-150, hep-th/9510129;\\
J.Minahan and D.Nemeschansky, Nucl.Phys., B468 (1996) 72-84, hep-th/9601059'\\
N.Dorey, V.Khoze and M.Mattis, Phys.Rev., D54 (1996) 7832-7848, hep-th/9607202; Nucl.Phys., B492
(1997) 607-622, hep-th/9611016

\bibitem{DVph}
F.Cachazo, K.A.Intriligator and C.Vafa, Nucl.Phys. B603 (2001) 3, hep-th/0103067;\\
F.Cachazo and C.Vafa, hep-th/0206017;\\
R.Dijkgraaf and C.Vafa, Nucl.Phys. B644 (2002) 3, hep-th/0206255; Nucl.Phys. B644 (2002) 21, hep-
th/0207106; hep-th/0208048;\\
L.Chekhov and A.Mironov, Phys.Lett. B552 (2003) 293, hep-th/0209085;\\
A.Klemm, M.Marino and S.Theisen, JHEP 0303 (2003) 051, hep-th/0211216;\\
H.Itoyama and A.Morozov, Nucl.Phys. B657 (2003) 53-78, hep-th/0211245; Phys.Lett. B555 (2003) 287-
295, hep-th/0211259; Prog.Theor.Phys. 109 (2003) 433-463, hep-th/0212032;
Int.J.Mod.Phys. A18 (2003) 5889-5906, hep-th/0301136; \\
L.Chekhov, A.Marshakov, A.Mironov and D.Vasiliev, hep-th/0301071; Proc. Steklov Inst.Math. 251 (2005)
254, hep-th/0506075; \\
M.Matone and L.Mazzucato, JHEP 0307 (2003) 015, hep-th/0305225; \\
R.Argurio, G.Ferretti and R.Heise, Int.J.Mod.Phys. A19 (2004) 2015-2078, hep-th/0311066; \\
M.Gomez-Reino, JHEP 0406 (2004) 051, hep-th/0405242; \\
H. Itoyama and H. Kanno, Nucl.Phys. B686 (2004) 155-164, hep-th/0312306; \\
H.Itoyama, K.Maruyoshi and M.Sakaguchi, Nucl.Phys. B794 (2008) 216-230, arXiv:0709.3166


\bibitem{CIV}
A.Morozov and Sh.Shakirov, arXiv:1004.2917

\bibitem{towaproof} A.Mironov, A.Morozov and Sh.Shakirov,
IJMP A27 (2012) 1230001, arXiv:1011.5629

\bibitem{AMM/EO}
A.Alexandrov, A.Mironov and A.Morozov, Int.J.Mod.Phys. A19 (2004) 4127, Theor.Math.Phys. 142 (2005)
349, hep-th/0310113; Int.J.Mod.Phys. A21 (2006) 2481, hep-th/0412099; Fortsch. Phys. 53 (2005) 512, hepth/
0412205; \\
B.Eynard, JHEP 0411 (2004) 031, hep-th/0407261; \\
B.Eynard and N.Orantin, JHEP 0612 (2006) 026, math-ph/0504058; math-phys/0702045;\\
L.Chekhov, B.Eynard, JHEP 0603 (2006) 014, hep-th/0504116; JHEP 0612 (2006) 026, math-ph/0604014;\\
B.Eynard, M.Marino and N.Orantin, JHEP 0706 (2007) 058, hep-th/0702110;\\
N.Orantin, PhD thesis, arXiv:0709.2992; arXiv:0803.0705; \\
R. Dijkgraaf and H. Fuji, Fortsch.Phys. 57 (2009) 825-856, arXiv:0903.2084;\\
I.Kostov and N.Orantin, arXiv:1006.2028;\\
L.Chekhov, B.Eynard and O.Marchal, arXiv:1009.6007;\\
R.Dijkgraaf, H.Fuji and M.Manabe, arXiv:1010.4542;\\
V.Bouchard and P.Sulkowski, arXiv:1105.2052; \\
S.Gukov and P.Sulkowski, arXiv:1108.0002

\bibitem{HZ} J. Harer, D. Zagier, Inv. Math. 85 (1986) 457-485; \\
S.K. Lando and A.K. Zvonkine, Graphs on Surfaces and Their Applications, Springer (2003);\\
E.Akhmedov and Sh.Shakirov, arXiv:0712.2448;\\
Sh.Shakirov and A.Morozov, JHEP 0912 (2009) 003, arXiv:0906.0036

\bibitem{BH} E.Brezin and S.Hikami,
JHEP 0710(2007)096, arXiv:0709.3378; Commun.Math.Phys.283(2008)507-521,
arXiv:0708.2210; arXiv:cond-mat/9804024

\bibitem{HZBH} A.Morozov and Sh.Shakirov, arXiv:1007.4100

\bibitem{virco}
R. Dijkgraaf, H. L. Verlinde, and E. P. Verlinde,  Nucl. Phys. B348 (1991) 435–456; \\
M. Fukuma, H. Kawai, and R. Nakayama, Int. J. Mod. Phys. A6 (1991) 1385–1406;\\
A. Mironov and A. Morozov, Phys.Lett. B252 (1990) 47-52;\\
F. David, Mod.Phys.Lett. A5 (1990) 1019;\\
J. Ambjorn and Yu. Makeenko, Mod.Phys.Lett. A5 (1990) 1753;\\
H. Itoyama and Y. Matsuo, Phys.Lett. 255B (1991) 202

\bibitem{GMMMO} A. Gerasimov, A. Marshakov, A. Mironov, A. Morozov, and A. Orlov,
Nucl. Phys. B357 (1991) 565-618

\bibitem{Wreps} A.Morozov and Sh.Shakirov, JHEP 0904:064,2009, arXiv:0902.2627;\\
A.Alexandrov, arXiv:1005.5715, arXiv:1009.4887;\\
A.Balantekin, arXiv:1011.3859

\bibitem{unimo} A.Morozov, Theor.Math.Phys.162:1-33,2010, arXiv:0906.3518

\bibitem{betadefo} A.Morozov, arXiv:1201.4595

\bibitem{GKM}
M. Kontsevich, Funkts. Anal. Prilozh., 25:2 (1991) 50-57; Comm.Math.Phys. 147 (1992) 1-23; \\
S.Kharchev, A.Marshakov, A.Mironov, A.Morozov and A.Zabrodin, Phys. Lett. B275 (1992)
311-314, hep-th/9111037; Nucl.Phys. B380 (1992) 181-240, hep-th/9201013; \\
P.Di Francesco, C.Itzykson and J.-B.Zuber, Comm.Math.Phys. 151 (1993) 193-219, hep-th/9206090

\bibitem{dua}
L.Chekhov and Yu.Makeenko, Phys.Lett. B278 (1992) 271-278, arXiv:hep-th/9202006; \\
S.Kharchev and A.Marshakov, hep-th/9210072;
S.Kharchev and A.Marshakov, Int. J. Mod. Phys. A 10 (1995) 1219, hep-th/9303100;\\
A.Mironov, A.Morozov and G.W.Semenoff, Int. J. Mod. Phys. A 11 (1996) 5031, hep-th/9404005;\\
A.Alexandrov, A.Mironov and A.Morozov, JHEP 0912(2009)053, arXiv:0906.3305

\bibitem{Okou} A.Okounkov, arXiv:math/0101201; \\
A.Alexandrov, A.Mironov, A.Morozov, P.Putrov, Int.J.Mod.Phys.A24 (2009) 4939-4998, arXiv:0811.2825

\bibitem{nal}
I.Gelfand, M.Kapranov and A.Zelevinsky, {\it Discriminants, Resultants and Multidimensional Determinants},
Birkhauser, 1994; \\
V.Dolotin, alg-geom/9511010;\\
V.Dolotin and A.Morozov, {\it Inroduction to Non-linear Algebra},
World Scientific, 2007, hep-th/0609022; \\
A.Morozov and Sh.Shakirov, arXiv:0911.5278; \\
K.Fujii, SIGMA 7 (2011), 022, arXiv:0912.2135; K.Fujii, H.Oike, arXiv:1103.4428;\\
U.Svensson, arXiv:0912.3172;\\
A.Stoyanovsky, arXiv:1103.0514;\\
J.Lasserre, arXiv:1110.6632

\bibitem{Mth}
A.Alexandrov, A.Mironov and A.Morozov,
Teor.Mat.Fiz.150:179-192,2007, hep-th/0605171;
Physica D235:126-167,2007, hep-th/0608228;
JHEP 0912:053,2009, arXiv:0906.3305

\bibitem{BGW}
E.Brezin and D.Gross, Phys.Lett., B97 (1980) 120; \\
D.Gross and E.Witten, Phys.Rev., D21 (1980) 446-453


\bibitem{Hut}
R.Dijkgraaf, In: {\it The moduli spaces of curves, Progress in Math.}, 129 (1995), 149-163, Birkhauser;\\
S.Lando and D.Zvonkine, Funk.Anal.Appl. 33 3 (1999) 178-188; math.AG/0303218; \\
A.Givental, math/0108100; \\
M.Kazarian, arXiv:0809.3263; \\
V.Bouchard and M.Marino, arXiv:0709.1458; \\
A.Mironov and A.Morozov, JHEP 0902 (2009) 024, arXiv:0807.2843; \\
G.Borot, B.Eynard, M.Mulase and B.Safnuk,   arXiv:0906.1206;\\
A.Morozov and Sh.Shakirov, Mod.Phys.Lett.A24:2659-2666,2009, arXiv:0906.2573

\bibitem{MMN}
A.Morozov, A.Mironov, S.Natanzon,
Theor.Math.Phys.166:1-22,2011, arXiv:0904.4227; JHEP 11 (2011) 097, arXiv:1108.0885; \\
A.Alexandrov, A.Morozov, A.Mironov and S.Natanzon, J.Phys. A45 (2012) 045209,
arXiv:1103.4100

\bibitem{AGT}
L.Alday, D.Gaiotto and Y.Tachikawa, Lett.Math.Phys. 91 (2010) 167-197, arXiv:0906.3219; \\
N.Wyllard, JHEP 0911 (2009) 002, arXiv:0907.2189;\\
N.Drukker, D.Morrison and T.Okuda, JHEP 0909 (2009) 031, arXiv:0907.2593;\\
A.Mironov and A.Morozov, Nucl.Phys. B825 (2009) 1-37, arXiv:0908.2569; Phys.Lett.
B680 (2009) 188-194, arXiv:0908.2190; \\
A.Mironov, S.Mironov, A.Morozov and And.Morozov, arXiv:0908.2064; \\
L.Hadasz, Z.Jaskolski and P.Suchanek, arXiv:0911.2353; arXiv:1004.1841; \\
S.Yanagida, arXiv:1005.0216; \\
V.Alba, V.Fateev, A.Litvinov and G.Tarnopolsky, arXiv:1012.1312

\bibitem{Nekf}
N.Nekrasov, Adv.Theor.Math.Phys. 7 (2004) 831-864, hep-th/0206161; \\
N.Nekrasov and A.Okounkov, hep-th/0306238

\bibitem{LMNS}
G.Moore, N.Nekrasov, S.Shatashvili, Nucl.Phys. B534 (1998) 549-611, hep-th/9711108;
hep-th/9801061; \\
A.Losev, N.Nekrasov and S.Shatashvili, Commun.Math.Phys. 209 (2000) 97-121,
hep-th/9712241; ibid. 77-95, hep-th/9803265

\bibitem{intSW}
A.Gorsky, I.Krichever, A.Marshakov, A.Mironov, A.Morozov, Phys.Lett. B355 (1995) 466-477,
hep-th/9505035; \\
E.Martinec, Phys.Lett., B367 (1996) 91-96;\\
E.Martinec and N.Warner, Nucl.Phys., 459 (1996) 97, hep-th/9509161; \\
T. Nakatsu and K. Takasaki, Mod. Phys. Lett. A 11 (1996) 157, hep-th/9509162;\\
R.Donagi and E.Witten, Nucl.Phys., B460 (1996) 299-334, hep-th/9510101;\\
A.Gorsky, S.Gukov and A.Mironov, Nucl.Phys., B517 (1998) 409-461; Nucl.Phys., B518
(1998) 689;\\
N.Nekrasov and S.Shatashvili, Nucl.Phys. Proc.Suppl. B192-193 (2009) 91-112,
arXiv:0901.4744; arXiv:0901.4748 \\
A.Mironov and A.Morozov, JHEP 04 (2010) 040, arXiv:0910.5670; J.Phys. A43 (2010)
195401, arXiv: 0911.2396; \\
A.Popolitov, arXiv:1001.1407; \\
A.Mironov, A.Morozov and Z.Zakirova, arXiv:1202.6029

\bibitem{confmamo}
Vl.Dotsenko and V.Fateev, Nucl.Phys. B240 (1984) 312-348; \\
A.Marshakov, A.Mironov, and A.Morozov, Phys.Lett. B265 (1991) 99;\\
S.Kharchev, A.Marshakov, A.Mironov, A.Morozov and S.Pakuliak, Nucl.Phys. B404 (1993)
17-750, arXiv:hep-th/9208044;\\
R.Dijkgraaf and C.Vafa, arXiv:0909.2453; \\
A.Mironov, A.Morozov and Sh.Shakirov,
JHEP 1002:030,2010, . arXiv:0911.5721;
Int.J.Mod.Phys. A25:3173-3207,2010, arXiv:1001.0563;
JHEP 1103:102,2011, arXiv:1011.3481  \\
H.Itoyama, K.Maruyoshi and T.Oota, Prog.Theor.Phys. 123 (2010) 957-987,
arXiv:0911.4244; \\
H.Itoyama and T.Oota, arXiv:1003.2929; \\
A.Mironov, A.Morozov and And.Morozov, arXiv:1003.5752; \\
T.Eguchi and K.Maruyoshi, arXiv:0911.4797; arXiv:1006.0828; \\
A.Mironov, A.Morozov, A.Popolitov, Sh.Shakirov, arXiv:1103.5470

\bibitem{HS}
A.Mironov, A.Morozov and Sh.Shakirov, JHEP 1102:067,2011, arXiv:1012.3137; \\
A.Mironov, A.Morozov, Sh.Shakirov and A.Smirnov, Nucl.Phys.  B 855 (2012), pp. 128-151,
arXiv:1105.0948

\bibitem{knth}
J.W.Alexander, Trans.Amer.Math.Soc. 30 (2) (1928) 275?306;\\
J.H.Conway, Algebraic Properties, In: John Leech (ed.), {\it Computational Problems in Abstract
Algebra}, Proc. Conf. Oxford, 1967, Pergamon Press, Oxford-New York, 329-358, 1970;\\
V.F.R.Jones, Invent.Math. 72 (1983) 1 Bull.AMS 12 (1985) 103Ann.Math. 126 (1987) 335;\\
L.Kauffman,Topology 26 (1987) 395

\bibitem{CS}
S.-S.Chern and J.Simons, Ann.Math. 99 (1974) 48-69;\\
E.Witten, Comm.Math.Phys. 121 (1989) 351;\\
G.Moore and N.Seiberg, Phys.Lett. B220 (1989) 422;\\
V.Fock and Ya.I.Kogan, Mod.Phys.Lett. A5 (1990) 1365-1372;\\
R.Gopakumar and C.Vafa, Adv.Theor.Math.Phys. 3 (1999) 1415-1443, hep-th/9811131

\bibitem{extHOMFLY}
A.Mironov, A.Morozov and And.Morozov, arXiv:1203.0667; arXiv:1112.5754

\bibitem{HOMFLY}
P.Freyd, D.Yetter, J.Hoste, W.B.R.Lickorish, K.Millet, A.Ocneanu, Bull. AMS. 12 (1985) 239;\\
J.H.Przytycki and K.P.Traczyk, Kobe J. Math. 4 (1987) 115-139; \\
E.Guadagnini, M.Martellini and M.Mintchev, In Clausthal 1989, Proceedings, Quantum groups,
307-317; Phys.Lett. B235 (1990) 275;\\
N.Yu.Reshetikhin and V.G.Turaev, Comm. Math. Phys. 127 (1990) 1-26; \\
R.K.Kaul and T.R.Govindarajan, Nucl.Phys. B380 (1992) 293-336, hep-th/9111063; \\
P.Ramadevi, T.R.Govindarajan and R.K.Kaul, Nucl.Phys. B402 (1993) 548-566,
hep-th/9212110; Nucl.Phys. B422 (1994) 291-306, hep-th/9312215; \\
R.M.Kashaev, Mod.Phys.Lett. A9 (1994) 3757-3768, hep-th/9411147; q-alg/9504020;
Lett.Math.Phys. 39 (1997) 269-265, q-alg/9601025; \\
H.Ooguri and C.Vafa, Nucl.Phys. B577 (2000) 419-438, hep-th/9912123;\\
P.Ramadevi and T.Sarkar, Nucl.Phys. B600 (2001) 487-511, hep-th/0009188; \\
J.Labastida, M.Marino, Comm.Math.Phys. 217 (2001) 423-449, hep-th/0004196; math/010418; \\
J.M.F.Labastida, M.Marino and C.Vafa, JHEP 0011 (2000) 007, hep-th/0010102;\\
M.Marino and C.Vafa, arXiv:hep-th/0108064;\\
D.Galakhov, A.Mironov, A.Morozov, A.Smirnov, arXiv:1104.2589; \\
Zodinmawia and P.Ramadevi, arXiv:1107.3918; \\
A.Mironov, A.Morozov and Sh.Shakirov, arXiv:1203.0667

\bibitem{superpols}
S.Gukov, A.Schwarz and C.Vafa, Lett.Math.Phys. 74 (2005) 53-74, arXiv:hep-th/0412243;\\
N.M.Dunfield, S.Gukov and J.Rasmussen, Experimental Math. 15 (2006) 129-159, math/0505662; \\
E.Gorsky, arXiv:1003.0916;\\
M.Aganagic and Sh.Shakirov, arXiv: 1105.5117; \\
P.Dunin-Barkowski, A.Mironov, A.Morozov, A.Sleptsov and A.Smirnov, arXiv:1106.4305 v2;\\
N.Carqueville and D.Murfet, arXiv:1108.1081;\\
I.Cherednik, arXiv:1111.6195;\\
S.Gukov and M.Stosic, arXiv:1112.0030; \\
A.Oblomkov, J.Rasmussen and V.Shende, arXiv:1201.2115 (with an Appendix by Eugene Gorsky);\\
A.Mironov, A.Morozov, Sh.Shakirov and A.Sleptsov, arXiv:1201.3339; \\
H.Fuji, S.Gukov and P.Sulkowski, arXiv:1203.2182; \\
H.Itoyama, A.Mironov, A.Morozov and And.Morozov, arXiv:1204.0913

\bibitem{BEM}
M.Rosso and V.F.R.Jones, J. Knot Theory Ramifications, 2 (1993) 97-112
X.-S.Lin and H.Zheng, Trans. Amer. Math. Soc. 362 (2010) 1-18 math/0601267;
S.Stevan, Annales Henri Poincar´e 11 (2010) 1201-1224, arXiv:1003.2861;
A.Brini, B.Eynard and M.Mari˜no, arXiv:1105.2012

\bibitem{ABJM}
O.Aharony, O.Bergman, D.L.Jafferis and J.Maldacena,   arXiv:0806.1218

\bibitem{ABJMsol}
A.Kapustin, B.Willett and I.Yaakov, JHEP 1003, 089 (2010), arXiv:0909.4559;\\
M.Marino and P.Putrov, JHEP 1006, 011 (2010), arXiv:0912.3074; \\
N.Drukker, M.Marino and P. Putrov,
arXiv:1007.3837; arXiv:1103.4844; \\
M.Marino,  arXiv:1104.0783; \\
H.Fuji, S.Hirano and S.Moriyama,  JHEP 1108:001, (2011), arXiv:1106.4631



\end{thebibliography}
\end{document}